\documentstyle[12pt,epsf]{article}

\def\be{\begin{equation}}
\def\ee{\end{equation}}
\def\bea{\begin{eqnarray}}
\def\eea{\end{eqnarray}}
\def\nn{\nonumber}

\newcommand{\Section}[1]{\section{#1}\setcounter{equation}{0}}

\begin{document}

\begin{flushright}
hep-th/0202025 \\
TIFR/TH/02-03
\end{flushright}

\pagestyle{plain} 

\def\e{{\rm e}}
\def\cs{\frac{1}{(2\pi\alpha')^2}}
\def\CV{{\cal{V}}}
\def\haf{{\frac{1}{2}}}
\def\tr{{\rm Tr}}
\def\"{\prime\prime}
\def\p{\partial}
\def\tphi{\tilde{\phi}}
\def\ttheta{\tilde{\theta}}
\def\hj{\hat j}
\def\hn{\hat n}
\def\bz{\bar{z}}
\def\zk{{\bf{Z}}_k}
\def\h1{\hspace{1cm}}
\def\dd{\Delta_{[N+2k] \times [2k]}}
\def\ddbar{\bar{\Delta}_{[2k] \times [N+2k]}}
\def\u{U_{[N+2k] \times [N]}}
\def\ubar{\bar{U}_{[N] \times [N+2k]}}
\def\goes{\rightarrow}
\def\goal{\alpha'\rightarrow 0}

\vspace{3cm}

\begin{center}
{\Large {\bf Non-Commutative Instantons and the Information Metric
}}
\end{center}

\vspace{.15cm}

\begin{center}
Shahrokh Parvizi
\footnote{On leave from {\it ``Institute for Studies in
Theoretical Physics and Mathematics"}, Tehran, Iran.}

\vspace{.3 cm}

{\it Department of Theoretical Physics,\\
Tata Institute of Fundamental Research,\\
Homi Bhabha Road, Mumbai 400 005, INDIA.}\\

\vskip .1 in 
%

{\sl parvizi@theory.tifr.res.in} 
\vspace{1cm} 
\end{center} 
\vskip .02 in
\begin{abstract} 
By using the so-called Information Metric on the moduli space of an
anti-selfdual (ASD) Instanton in a Self-Dual (SD) Non-Commutative
background, we investigate the geometry of moduli space. The metric is
evaluated perturbatively in non-commutativity parameter and we show that
by putting a cut-off in the location of instanton in the definition of
Information Metric we can recover the five dimensional space time in the
presence of a B-field. This result shows that the non-commutative
YM-Instanton Moduli corresponds to D-Instanton Moduli in the gravity side
where the radial and transverse location of D-Instanton are corresponding
to YM-Instanton size and location, respectively. The match is shown in the
first order of non-commutativity parameter.

\end{abstract}

\vspace{3cm}
{\small 
\begin{tabular}{l l}
KEYWORDS: &Instanton, Non-Commutative Gauge Theory, Moduli Space, \\
 &AdS/CFT Correspondence.\\
PACS Numbers: & 11.15.-q, 04.65.+e, 02.40.Gh\\

\end{tabular}
}
\newpage
\Section{Introduction}

In the study of AdS/CFT correspondence \cite{oz}, the instantons are of
particular interest as a non-perturbative probe. They entered in the
subject through the work of Dorey, {\it et al} works \cite{dorey}, where
they showed that, the Yang-Mills multi-instanton measure is related to the
D-instanton counterpart in the gravity theory of the bulk. To reach this
correspondence, the saddle point approximation is used for $SU(N)$
Yang-Mills in the large $N$ limit and the result is that the integral
measure of YM instantons is equivalent to the D(-1)-brane (D-Instanton)
effective action in the bulk. Moreover in the large $N$ and large $k$
limit ($k$ is the instantons number), the effective action corresponds to
the $U(k)$ matrix model in IIB theory which, in turn is the D-Instanton
effective action.

Since for $SU(N)$ gauge theory, every instanton lives in a $SU(2)$
subgroup, actually, the integration over zero modes of the gauge
orientations in $SU(N)$ gives effectively the moduli of the instanton in
$SU(2)$. Beside following the method of integration over zero modes and
saddle point approximation for large $N$ limit, it is worth to ask for a
direct relation between the moduli space of YM instantons on the boundary
and the moduli space of D-instantons on the bulk. 

However, we know that the moduli of the D-instanton is the $AdS_5$ itself,
but it is not so clear for the YM instanton moduli. The most natural
metric for the YM instanton moduli is the usual $L^2$ metric in the field
configuration space. However, this metric for a single instanton on ${\bf
R^4}$ is a flat metric and for an instanton on $S^4$ (which is the
$AdS_5$ boundary) is a very complicated metric \cite{moduli}. It
means with $L^2$ metric the direct map between two moduli spaces in gauge
theory and gravity sides is impossible. Therefore, we need a new
definition for the metric on the moduli space.

Fortunately, as proposed by Blau, {\it et al}\cite{narain}, there exists
an interesting metric on the YM instantons moduli called Information
Metric (formerly introduced in \cite{hitchin}) which can do the job. This
metric is constructed simply by the Instanton density $\tr_N F^2$ (see
(\ref{3.05})) and has the important property of conformal invariance which
makes it a good candidate metric for the boundary theory.

It has been shown in \cite{narain}, that the Information Metric for a
single instanton on ${\bf R^4}$ (and also on $S^4$) is an $AdS_5$ metric
in which the radial coordinate corresponds to the instanton size and the
other transverse coordinates indicate the location of instanton in ${\bf
R^4}$. Therefore, this is the exact correspondence between YM instanton
moduli space and the D-instanton location in the Supergravity Theory.
Moreover, in \cite{narain}, the authors consider possible small
fluctuations of the base geometry around ${\bf R^4}$ and find the
correction to the $AdS_5$ metric. In this direction, one can think about
another deformation of ${\bf R^4}$, particularly non-commutative 
${\bf R^4_{NC}}$, and look for the corresponding instanton moduli and
compare it with the Supergravity side.

On the gravity side, it is worth mentioning the work of Maldacena and
Russo \cite{maldrusso} in which the AdS/CFT correspondence is understood
as correspondence of
the non-commutative YM theories in the boundary to Supergravity with
anti-symmetric $B$ field in the bulk. The latter has an Einstein metric
(in selfdual $B$ background \cite{das}) and contains the D-instantons. The
non-commutativity is related to the boundary value of $1/B$. So in this
context, it is quiet natural to expect that an extension of YM instantons
to the ${\bf R^4_{NC}}$ background may have a correspondence to the
Supergravity D-instantons in the Maldacena-Russo background.

On the other hand, in the gauge theory side, the non-commutative
instantons are also well-known. The general prescription, as a
non-commutative version of ADHM construction \cite{ADHM}, introduced in
the work by Nekrasov and Schwarz in \cite{nekrasov} and the more explicit
solutions for the non-commutative ADHM equations are discussed in several
works \cite{furuuchi,lee,kim1,correa,kim,chu,hamanaka,popov,ishikawa}.

The purpose of this letter is to extend the idea of the Information Metric
to the NC single $SU(N)$ instanton and find the relation with the
D-instanton in \cite{maldrusso}, if any.

The extension needs a careful definition for the Information Metric in
the NC background. To overcome some technical problems and to recover
the commutative case, we study the Information Metric in the small
non-commutativity limit. By proposing a modified definition for the
Information Metric, we can relate it to the Supergravity metric in the 
first order of approximation. 

\Section{Non-Commutative Instantons}

In the following subsections, the instanton construction in a
non-commutative space is reviewed along which we introduce our
notations.

\vspace{10mm} 

\noindent {\large{\it 2.1 Non-Commutative ${\bf R^4}$}}

Let us start by the non-commutative Euclidean four dimensional space,
${\bf R^4_{NC}}$. In general, the non-commutativity is defined by the
commutator of coordinates as follows,
\bea\label{2.10}
[x^\mu, x^\nu]= i \theta^{\mu\nu}
\eea
where $\theta$ is a real anti-symmetric constant tensor.
If we consider $\theta$ to be a rank 4 tensor, then by four dimensional
rotations and re-scaling, it is possible to transform it to the following
form,
\bea \label{2.15}
\theta^{12}=\pm \theta^{34}
\eea 
with the other components equal to zero. In (\ref{2.15}), the $+$ and
$-$ signs define self-dual and anti-selfdual $\theta$ tensor,
respectively. We use self-dual $\theta$ in this letter and set $\theta^{12}=
\theta^{34}\equiv \zeta /2$. \footnote{It can be converted to the
anti-self-dual case by a parity transformation.} 

It is also convenient to use complex coordinates defined as: 
\bea \label{2.20}
z_1 = x_2 + i x_1, \;\; \; \bz_1 = x_2 - i x_1,\\
z_2 = x_4 + i x_3, \;\; \; \bz_2 = x_4 - i x_3 \;.
\eea
Then, the commutation relations will be 
\bea \label{2.25}
[z_i, \bz_j] = -\zeta\delta_{ij} 
\eea

The above commutators have a harmonic oscillator representation in which 
coordinates $z$ act as operators on the basis $|n_1, n_2\rangle$ as
follows,
\bea\label{2.30}
z_1\; |n_1, n_2\rangle = \sqrt{\zeta}\sqrt{n_1+1}\;|n_1+1,
n_2\rangle,\;\;\;
\bz_1\; |n_1, n_2\rangle = \sqrt{\zeta}\sqrt{n_1}\;|n_1-1, n_2\rangle,\\  
z_2\; |n_1, n_2\rangle = \sqrt{\zeta}\sqrt{n_2+1}\;|n_1,
n_2+1\rangle,\;\;\;  
\bz_2\; |n_1, n_2\rangle = \sqrt{\zeta}\sqrt{n_2}\;|n_1, n_2-1\rangle,  
\eea
and as usual:
\bea \label{2.35}
z_i \bz_i\; |n_1, n_2\rangle = \zeta n_i\; |n_1, n_2\rangle, \;\;\; (i=1,
2\;\; {\rm no \;\;sum}).
\eea

The relation between operator valued functions and ordinary c-functions
can be obtained by the following map:
\bea\label{2.36}
\Omega\; : \hat{\phi}(\hat{z},\hat{\bar{z}})\goes
\tilde{\phi}(z,\bar{z})=\langle 
z|\hat{\phi}(\hat{z},\hat{\bar{z}})|\bar{z}\rangle,
\eea
where $|\bar{z}\rangle$ and $\langle z|$ are the normalized coherent
states,
\bea\label{2.37}
\hat{\bz}|\bz\rangle=\bz|\bz\rangle, \;\;\; \langle z| \hat{z} = \langle 
z| z, \;\; \langle z|\bz \rangle=1 \; .
\eea
In this map, one can also relate sum over states to an integral over
functions as follows,
\bea\label{2.39}
\Omega\; : (\zeta\pi)^2 {\rm
Tr}_{\cal H}\hat{\phi}(\hat{z},\hat{\bar{z}})\goes
\int d^4z \tilde{\phi}(z,\bar{z})
\eea
where ${\cal H}$ is the representing Hilbert space.

\vspace{10mm}

\noindent {\large{\it 2.2 The ADHM Construction of Instantons}}

SD or ASD instantons are defined as field configurations for SD or ASD
Yang-Mills field, respectively, and subject to the constraint,
\bea\label{2.40}
Q := - \frac{1}{16\pi^2} \int d^4x\;  {\rm Tr}_N F\wedge F = \pm k
\eea
where $Q$ is known as the instanton charge and its absolute value is an
integer constant $k$. The sign $\pm$ corresponds to SD or ASD instantons, 
respectively.

The solutions can be found by an algebraic ansatz so-called  ADHM
construction \cite{ADHM}. Here, we briefly review the construction. 

Firstly, consider an $(N+2k) \times (2k)$
matrix $\dd$ which is linear in coordinates as follows,
\bea\label{2.45}
\dd &=& a_{[N+2k]\times [2k]} + b_{[N+2k]\times [2k]} x_{[2]\times
[2]} \h1 {\rm for\; SD \; Instantons} \nn\\
\dd &=& a_{[N+2k]\times [2k]} + b_{[N+2k]\times [2k]} \bar{x}_{[2]\times
[2]} \h1 {\rm for\; ASD \; Instantons}
\eea     
where $a$ and $b$ are constant complex valued matrices and 
we have used quaternionic representation for coordinates $x_{\mu}$,
\bea\label{2.47}
\bar{x} = \left(\matrix{-\bz_1 & z_2\cr   
  -\bz_2 &  - z_1} \right)\;, \h1\h1
x = \left(\matrix{-z_1 & -z_2\cr
   \bz_2 &  - \bz_1} \right)
\eea

Since $\dd$ has $N$ more rows than columns, it has an $N$ dimensional null
space. We put independent null vectors of $\bar{\Delta}$ in an  
$(N+2k) \times N$ dimensional matrix $U(x)$, {\it i.e.}, 
\bea\label{2.50}
\ddbar \u &=& 0 \\
\ubar \dd &=& 0
\eea 
with the normalization condition, 
\bea\label{2.55}
\ubar \u = 1_{[N]\times [N]}.
\eea
Then the gauge potential in this ansatz will be derived from $U$ as
follows, 
\bea\label{2.60}
A_{\mu [N]\times[N]} = \ubar \p_\mu \u
\eea

To have an ASD or SD $F_{\mu\nu}$ for non-zero $k$, we need two
additional conditions. 
\bea\label{2.65}
\ddbar \dd = 1_{[2]\times [2]} f^{-1}_{[k]\times[k]} 
\eea
\bea\label{2.67}
\dd f_{[k]\times[k]} \ddbar + \u \ubar = 1_{[N+2k]\times [N+2k]}
\eea
where $f$ is an arbitrary $x$-dependent $k \times k$ matrix. Using the
above relations, it is easy to write the instanton field strength as,
\bea \label{2.70}
F^{ASD}_{\mu\nu} &=& 4 \bar{U} b \bar{\sigma}_{\mu\nu}f\bar{b} U\\
F^{SD}_{\mu\nu} &=& 4 \bar{U} b \sigma_{\mu\nu}f\bar{b} U
\eea
where 
\bea \label{2.75}
\bar{\sigma}_{\mu\nu} = \frac{1}{2} i \bar{\eta}^a_{\mu\nu} \tau^a
\h1 {\rm and} \h1
\sigma_{\mu\nu} = \frac{1}{2} i \eta^a_{\mu\nu} \tau^a
\eea
with $\eta^a_{\mu\nu}$'s the standard 'tHooft symbols and $\tau^a$
the Pauli matrices. The anti-selfduality (selfduality) of $F_{\mu\nu}$
comes simply from anti-selfduality (selfduality) of
$\bar{\sigma}_{\mu\nu} (\sigma_{\mu\nu})$, respectively. 

The above construction is valid for non-commutative case too. In the
non-commutative case we must care about the coordinates which are
operators. To be more specific, we convert $\bar{\Delta}$ to the
following matrix form,
\bea\label{2.77}
\bar{\Delta} \equiv \left(\matrix{I & B^{\dagger}_1 - \bz_1 & -B_2 + z_2
\cr
   J^{\dagger} & B^{\dagger}_2 - \bz_2 & B_1 - z_1} \right) 
\eea
where $I_{[k]\times [N+k]}$, $J_{[N+k]\times [k]}$ and $B_{[k]\times
[k]}$ are constant matrices including in the matrix $a_{[N+2k]\times
[2k]}$ in (\ref{2.45}). Then the condition
(\ref{2.65}) is equivalent to the following equations known as ADHM
equations in the non-commutative space \cite{nekrasov},
\bea\label{2.78}
[B_1, B^{\dagger}_1] + [B_2, B^{\dagger}_2] + I I^{\dagger} - J^{\dagger}
J = [z_1, \bz_1] + [z_2, \bz_2] = -\zeta 
\eea
\bea\label{2.79}
[B_1, B^{\dagger}_2] + I J^{\dagger} = 0
\eea

\vspace{10mm}

\noindent {\large{\it 2.3 $k=1$ Instantons}}

The main subject in the ADHM ansatz is to construct $\Delta$ and $U$
such that the conditions (\ref{2.65}) and (\ref{2.67}) are
satisfied. Even in the commutative case, it is difficult to solve the
problem for an arbitrary $k$. Here we recall the final solution for
one instanton ($k=1$) in the commutative ${\bf R^4}$:
\bea \label{2.80}
F^{\mu\nu}= -4 \eta^a_{\mu\nu} \frac{\rho^2}{[(x-a)^2+\rho^2]^2} 
\eea
\bea\label{2.82}
{\rm Tr}_N F^2 = 96 \frac{\rho^4}{[(x-a)^2+\rho^2]^4}
\eea
where $\rho$ and $a^{\mu}$ are the instanton size and location,
respectively. These parameters are usually encoded in the matrix 
$a_{[N+2k]\times [2k]}$.
 
For the non-commutative ${\bf R^4_{NC}}$ and $k=1$, we look at the ASD
instanton in the SD non-commutative background\footnote{The moduli of ASD
instantons on ASD background is equivalent to the commutative case, since
the instanton equations in both commutative cases and non-commutative 
cases are the same by the Seiberg-Witten map \cite{SW}.} and according
to \cite{kim} (see also \cite{chu}) the result will be as follows,
\bea\label{2.85}
{\rm Tr}_N F^2 = S_1 p + S_2
\eea
where the instanton is located at the origin and 
\bea\label{2.87}
S_1 &:=& 4 \left( (C_1^2 +C_4^2) z^2 (z^2 + \zeta)+2 C_2^2 (z^2
-\frac{\zeta}{2})z^2  \right) \nn\\
S_2 &:=& 8 C_3^2 (z^2 +\zeta)(z^2 +\frac{3\zeta}{2}). 
\eea
with
\bea\label{2.90}
C_1&:=& \frac{-2(\rho^2 +\zeta)}{z^2(z^2 + \rho^2 + \frac{\zeta}{2})(z^2
+ \rho^2 + \zeta)}\nn\\
C_2&:=& \frac{2\rho\sqrt{\rho^2 +\zeta}}{z^2(z^2 + \rho^2 +
\frac{\zeta}{2})(z^2 + \rho^2 + \zeta)}\left(\frac{z^2 + \rho^2 +
\zeta}{z^2 + \rho^2}\right)^{\frac{1}{2}} \nn\\
C_3&:=& \frac{2\rho\sqrt{\rho^2 +\zeta}}{(z^2+\zeta)(z^2 + \rho^2 +
\frac{3\zeta}{2})(z^2 + \rho^2 + \zeta)}\left(\frac{z^2 + \rho^2 +
\zeta}{z^2 + \rho^2 + 2\zeta}\right)^{\frac{1}{2}} \nn\\
C_4&:=& \frac{-2\rho^2}{(z^2+\zeta)(z^2 + \rho^2 +
\frac{3\zeta}{2})(z^2 + \rho^2 + \zeta)}\nn\\
z^2 &:=& z_1\bz_1 +z_2 \bz_2 \nn
\eea
and $p$ is a projection operator which excludes $|0, 0\rangle$ state. The
matrix
representation of $F^2$ in $|n_1, n_2\rangle$ basis is obtained simply by
replacing $z^2$ with $\zeta n/2\equiv \zeta/2 (n_1+n_2)$ and it will be
diagonal as expected from the radial symmetry of $F^2$ in four dimensional
space ${\bf R^4}$.

It can be shown that the instanton charge for this case is indeed exactly
$-1$, \cite{kim,chu},
\bea\label{2.95}
Q=\frac{-1}{16\pi^2}\int d^4z {\rm Tr}_N F^2(z^2) &=&
\frac{-1}{16\pi^2}(\frac{\zeta\pi}{2})^2\sum_{n_1,n_2}\langle n_1,
n_2|{\rm Tr}_N \hat{F}^2|n_1, n_2\rangle \nn\\
 &=& \frac{-\zeta^2}{64}\sum_{n}(n+1) F^2(\zeta n/2) = -1
\eea
where $F^2(\zeta n/2)$ is the matrix element of ${\rm Tr}_N \hat{F}^2$.


\Section{The Information Metric}

As stated in the introduction, the Information Metric is an alternative to
construct the geometry of the instantons moduli space. It can be defined
in terms of the field density as follows,
\bea\label{3.05}
G_{AB}& &:= \frac{5}{256\pi^2}\int d^4 z F^2 \p_A\log F^2 \p_B\log F^2\\
& &= \frac{5}{256\pi^2}\int d^4 z \frac{\p_A F^2 \p_B F^2}{F^2}
\eea
where $F^2 \equiv {\rm Tr}_N F^2$ and $\p_A \equiv \p/\p y^A$ in which
$y^A$ are the moduli space parameters and correspond to the location and
size of instantons, $y^A= (a^{\mu}, \rho)$.

Firstly, let us quote from \cite{narain}, the resulting metric for the
commutative single instanton  (\ref{2.80}),
\bea\label{3.10}
ds^2 =  \frac{1}{\rho^2}\left(d\vec{a}^2+d\rho^2\right)
\eea
The above metric is the well-known Euclidean $\rm AdS_5$ metric. This
result shows that the moduli space of YM-Instanton equipped with the
Information Metric has exactly the bulk geometry in the AdS/CFT
context. This is the most direct correspondence between YM-Instantons and
D-Instantons. In the following, we consider possible definitions of
the Information Metric for the NC YM-Instantons to explore this
correspondence more.

The natural extension of the Information Metric to a
non-commutative instanton, is to replace the commutative $F^2$ in
(\ref{3.05}) by its operator version in non-commutative case
(\ref{2.85}). The integration is also replaced by a Trace over Hilbert
Space of non-commutative space. So, we start with the following
definition:
\bea \label{3.15}
G_{AB} := \frac{5}{256\pi^2} (\frac{\zeta\pi}{2})^2 {\rm Tr}_{\cal H}
\hat{F}^2 \p_A\log \hat{F}^2 \p_B\log \hat{F}^2
\eea
or
\bea\label{3.20}
G_{AB} := \frac{5}{256\pi^2} (\frac{\zeta\pi}{2})^2 {\rm Tr}_{\cal H}
\frac{1}{\hat{F}^2}\p_A \hat{F}^2 \p_B \hat{F}^2 \;\; .
\eea
The first difficulty arises here with the ambiguity in ordering of the
terms. We can use the symmetry
between the indices $A$ and $B$ in the metric to fix the position of $\p_A
F^2$ and $\p_B F^2$ by symmetrizing them. The position of
$\frac{1}{F^2}$ is still ambiguous, and we can consider either of two
cases bellow,
\bea\label{3.22}
G_{AB}^{(1)} &:=& \frac{5}{256\pi^2} (\frac{\zeta\pi}{2})^2 {\rm
Tr}_{\cal
H} \frac{1}{\hat{F}^2}\p_{\{A} \hat{F}^2 \p_{B\}} \hat{F}^2 \\
G_{AB}^{(2)} &:=& \frac{5}{256\pi^2} (\frac{\zeta\pi}{2})^2 {\rm
Tr}_{\cal H} \p_{\{A} \hat{F}^2 \frac{1}{\hat{F}^2}\p_{B\}}
\hat{F}^2\;\; . 
\eea
However, it can be shown that for the radially symmetric class of solutions
(including (\ref{2.85})), both $G_{AB}^{(1)}$ and $G_{AB}^{(2)}$ have the
same result. Moreover, the Information Metric is diagonal in this case. 

To calculate the Information Metric, firstly, we need the derivatives of
$F^2$. Here, we put the instanton at the point with complex coordinates
$(a_i,\bar{a}_i)$, now taking
\bea 
z_i&=&x_i-a_i \nn\\
\bz_i&=&\bar{x}_i-\bar{a}_i \;\;.\nn
\eea
Note that the non-commutative algebra (\ref{2.25}) is valid for both
$z$ and $x$ coordinates. Then we have the following
derivatives involved in the Information Metric: 
\bea\label{3.25}
\frac{\p \hat{F}^2}{\p a_i} &=& -\frac{\p \hat{F}^2}{\p z_i} = -
\frac{2}{\zeta} [\bz_i,\hat{F}^2] \nn\\
\frac{\p \hat{F}^2}{\p \bar{a}_i} &=& -\frac{\p \hat{F}^2}{\p
\bz_i} = \frac{2}{\zeta} [z_i,\hat{F}^2] \\
\frac{\p \hat{F}^2}{\p \rho} & &.\nn
\eea

We use the harmonic oscillator basis to find the matrix elements of terms
in (\ref{3.20}) as follows,
\bea\label{3.30}
\langle m_1,m_2|\frac{1}{\hat{F}^2}|n_1,n_2\rangle &=& \frac{1}{F^2(\zeta
n/2)}  \delta_{\vec{m}\vec{n}}\nn\\
\langle m_1,m_2|\frac{\p \hat{F}^2}{\p \rho}|n_1,n_2\rangle &=&
\frac{\p F^2(\zeta n/2)}{\p \rho} \delta_{\vec{m}\vec{n}}\nn\\
\langle m_1,m_2|\frac{\p \hat{F}^2}{\p a_1}|n_1,n_2\rangle &=&
\sqrt{\frac{2}{\zeta}} \left(W(n-1)\sqrt{n_1} \delta_{m_1+1,n_1}
\delta_{m_2n_2}-S_1(0)\delta_{\vec{m}0}\delta_{n_11}
\delta_{n_20}\right) \nn\\ 
\langle m_1,m_2|\frac{\p \hat{F}^2}{\p a_2}|n_1,n_2\rangle &=& 
\sqrt{\frac{2}{\zeta}} \left(W(n-1)\sqrt{n_2}
\delta_{m_2+1,n_2} 
\delta_{m_1n_1}-S_1(0)\delta_{\vec{m}0}\delta_{n_21}
\delta_{n_10}\right) \nn\\
\langle m_1,m_2|\frac{\p \hat{F}^2}{\p \bar{a}_1}|n_1,n_2\rangle &=&
\sqrt{\frac{2}{\zeta}}\left(W(n)\sqrt{n_1+1} \delta_{m_1-1,n_1} 
\delta_{m_2n_2}-S_1(0)\delta_{\vec{m}0}\delta_{n_11}
\delta_{n_20}\right) \nn\\
\langle m_1,m_2|\frac{\p \hat{F}^2}{\p \bar{a}_2}|n_1,n_2\rangle &=& 
\sqrt{\frac{2}{\zeta}}\left(W(n)\sqrt{n_2+1}
\delta_{m_2-1,n_2}
\delta_{m_1n_1}-S_1(0)\delta_{\vec{m}0}\delta_{n_10}
\delta_{n_21}\right) \nn\\
\eea
where $W(n)\equiv S_1(\zeta n/2)+ S_2(\zeta n/2)$ and $n=n_1+n_2$.

Now, one can write the Information Metric as follows,\footnote{As
mentioned before, we know that in the end of calculation
we will have $G_{AB}^{(1)}=G_{AB}^{(2)}\equiv G_{AB}$.}
\bea \label{3.35}
G_{AB} = \frac{5}{256\pi^2} (\frac{\zeta\pi}{2})^2 
\sum_{\vec{n},\vec{m},\vec{k}}^{\infty}
\langle \vec{n}|\frac{1}{F^2}|\vec{m}\rangle\langle \vec{m}|\p_{\{A} 
F^2|\vec{k}\rangle\langle \vec{k}|\p_{B\}} F^2|\vec{n}\rangle 
\eea

The sums over two indices, say $m$ and $k$, can be done easily by the help
of Kronecker delta functions in (\ref{3.30}) and the non-vanishing
components will be as follows,
\bea\label{3.40}
G_{\rho\rho}&=& \frac{5}{256}\frac{\zeta^2}{4}\left(\sum_{n_1,n_2\neq
(0,0)}^{\infty}
\frac{(\p W(n)/\p \rho)^2}{W(n)} + H_0 \right) \nn\\
&=&\frac{5}{256}\frac{\zeta^2}{4}\left(\sum_{n=1}^{\infty} (n+1) \frac{(\p
W(n)/\p \rho)^2}{W(n)} + H_0 \right)
\eea
\bea\label{3.45}
G_{a_i\bar{a}_i}&=&\frac{5}{256}\frac{\zeta^2}{4} 
\left(\sum_{n_1,n_2\neq(0,0)}^{\infty} \frac{2}{\zeta}
\frac{n_i(T(n-1))^2+(n_i+1)(T(n))^2}{2 W(n)} + G_0 
\right) \nn\\
&=&\frac{5}{256}\frac{\zeta^2}{4} \left(\sum_{n=1}^{\infty}
(n+1)\frac{2}{\zeta}\frac{n(T(n-1))^2+(n+2)(T(n))^2}{4 W(n)} +
G_0 \right) 
\eea
where,
\bea\label{3.47}
H_0 &=& \frac{(\p S_2(0)/\p \rho)^2}{S_2(0)} \\
G_0 &=&\frac{2}{\zeta}\left(\frac{(S_1(0))^2-2
T(0) S_1(0)}{2 W(1)} + \frac{(T(0)-S_1(0))^2}{2 S_2(0)}\right)
\eea
and $T(n)=W(n)-W(n+1)$. We have also used the following identities for
summations over $n_1$ and $n_2$,
\bea\label{3.50}
\sum_{n_1,n_2}^{\infty} g(n)n_1=\sum_{n_1,n_2}^{\infty}
g(n)n_2 = \sum_{n_1,n_2}^{\infty} g(n) n/2 = \sum_{n}^{\infty}
(n+1) g(n) n/2 \nn
\eea

In contrast to (\ref{2.95}), the resulting sum over $n$ for the
Information Metric is too complicated to have an
interesting interpretation.
Here, we consider small non-commutativity approximation. 
Firstly , for the limit $\zeta\goes 0$, the sums over $n$ go to an
integral. It can be seen for a continuous function $g$  as follows,
\bea\label{3.55}
\lim_{\zeta\goes 0} \sum_{n=1}^{\infty} (n+1) g(n\zeta/2)
 = \frac{2}{\zeta} \int_{\zeta/2}^{\infty} dt
(\frac{2t}{\zeta}+1) g(t) +{\cal O}(\zeta^2)
\eea
This approximation is valid up to the first order in $\zeta$. To check
this limit, we apply it to the instanton charge formula (\ref{2.95}),
\bea\label{3.60}
Q&=& \frac{-\zeta^2}{64}\sum_{n}(n+1) F^2(\zeta n/2) =
\frac{-\zeta^2}{64}\frac{2}{\zeta} \int_{\zeta/2}^{\infty} dt
(\frac{2t}{\zeta}+1) F^2(t) +{\cal O}(\zeta^2)\nn\\
&=&-1+{\cal O}(\zeta^2)
\eea

After converting the series in (\ref{3.40}) and (\ref{3.45}) to
integrals, the components of the Information Metric can be obtained by
expanding the integrands in a power series of $\zeta$ (while fixing 
$t$). The results are in the following,
\bea\label{3.65}
G_{\rho\rho}&=&\frac{5}{256}\frac{\zeta^2}{4}\left(\frac{2}{\zeta} 
\int_{\zeta/2}^{\infty} 
\left(
\frac{1536 (\rho^3-\rho t)^2}{(\rho^2+t)^6}-
\frac{9216 (\rho^3-\rho t)^2 \zeta
}{(\rho^2+t)^7}+{\cal O}(\zeta^2)  \right)(\frac{2t}{\zeta}+1)dt +
H_0 \right) \nn\\
&=& \frac{1}{\rho^2}\left(1-\frac{\zeta}{\rho^2} +{\cal
O}(\frac{\zeta^2}{\rho^4})\right)\\
G_{a_i\bar{a}_i}&=&\frac{5}{256}\frac{\zeta^2}{4}\left(\frac{2}{\zeta} 
\int_{\zeta/2}^{\infty}
\left(\frac{192 t \rho^4}{(\rho^2+t)^6} +
\frac{96 (2 t^2 \rho^2 - 9 t \rho^4 +\rho^6) \zeta
}{(\rho^2+t)^7}+{\cal O}(\zeta^2)  \right)(\frac{2t}{\zeta}+1)dt + 
G_0 \right)   \nn\\
&=& \frac{1}{2\rho^2}\left(1-\frac{\zeta}{2\rho^2} +
\frac{45\zeta}{64\rho^2}+{\cal O}(\frac{\zeta^2}{\rho^4})\right)
\eea
where the third term in the last equation is the contribution of
$G_0$.\footnote{The contribution of $H_0$ in (\ref{3.65}) is in the second
order of $\zeta$.} Finally, we have the following metric,
\bea\label{3.70}
ds^2 = \left(1+\frac{13\zeta}{64\rho^2} 
+{\cal O}(\frac{\zeta^2}{\rho^4})\right)\frac{d\vec{a}^2}{\rho^2} 
+\left(1-\frac{\zeta}{\rho^2} 
+{\cal O}(\frac{\zeta^2}{\rho^4})\right)\frac{d\rho^2}{\rho^2} 
\eea

Unfortunately, this metric does not agree with the expected results in
Supergravity side. Actually, if we neglect the contribution of $G_0$, it
will be in agreement with the SUGRA result (see section 4). This means
excluding the ground state $|0,0\rangle$ from $\tr_{{\cal H}}$ in the
definitions of Information Metric. In the coherent states language, the
$\tr_{{\cal H}}$ is replaced by integral over $z$ coordinates and the
exclusion of $|0,0\rangle$ state is equivalent to excluding the origin
from the
integration region. So we can use the following expression as Modified 
Information Metric,
\bea\label{3.72}
\tilde{G}_{AB} &=& \frac{5}{256\pi^2}\int_{{\cal M}'} d^4 z \langle z|
\frac{1}{\hat{F}^2}\p_{\{A}\hat{F}^2 \p_{B\}} \hat{F}^2|\bz\rangle \nn\\
 &=& \frac{5}{256\pi^2}\int_{{\cal M}'} d^4 z 
\widetilde{(\frac{1}{F^2})} *\widetilde{\p_{\{A} F^2} *\widetilde{\p_{B\}}
F^2}
\eea
where ${\cal M}' = {\bf R^4}-{\bf B^0}(\eta \zeta/2)$ and ${\bf
B^0}(\eta \zeta/2)$ is a ball centered at the origin with radius $\eta
\zeta/2$. $\tilde{F}^2$ shows the representation of $\hat{F}^2$ in the
coherent state formalism (see equation (\ref{2.36})). Using approximation
in the first order of
$\zeta$, the resulting metric will be as follows,\footnote{Note that the
cut-off $\eta$ will not appear in the first order of $\zeta$.} 
\bea\label{3.99} 
\widetilde{ds^2} &=& \frac{1}{\rho^2}
\left(1-\frac{\zeta}{\rho^2}
+{\cal O}(\frac{\zeta}{\rho^2})^2\right) d\vec{a}^2 \nn\\
& &+\frac{1}{\rho^2}\left(1-\frac{2\zeta}{\rho^2}
+{\cal O}(\frac{\zeta}{\rho^2})^2\right)d\rho^2
\eea

In the following section we will discussed the
above metric as a good candidate for the instanton metric which can be
matched with the Supergravity.

\Section{The AdS/CFT Correspondence}

In the context of String Theory, the non-commutativity can be traced in a
background which contains anti-symmetric $B$ field. Here, we consider an
Anti-Selfdual $B$ field extended in four directions and the following
solution will be derived from the Supergravity equations in Euclidean
Space-Time, \cite{maldacena, das}
\bea\label{4.00}
ds^2 &=&
\frac{\alpha'R^2}{\sqrt{\hat{g}}}\left[(f(u))^{-1/2}(dx_0^2+\cdots+dx_3^2)+ 
(f(u))^{1/2} (du^2+u^2d\Omega_5^2)\right] \\
B_{01}&=&B_{23}=-\frac{i\alpha' a^2 R^2}{\hat{g}}(f(u))^{-1}
\eea
where $R$ is the radius of ${\rm AdS}_5$, $\hat{g}$ is the normalized
String coupling and 
\bea\label{4.02}
f(u)=a^4+\frac{1}{u^4} .
\eea

The non-commutativity can be introduced as the parameter $a^2$ which is
the inverse of $B$ in the limit $u\goes \infty$. So we take 
$a^2 =\lambda \zeta$, where $\lambda$ is a constant to control the $B$
field and $\zeta$ independently. Firstly, let us change the coordinate
$u$ to $w=1/u$ which the latter has the length dimension,
\bea\label{4.04}
ds^2 =
\frac{\alpha'R^2}{\sqrt{\hat{g}}}\left[(1+a^4 
w^4)^{-1/2}\frac{d \vec{x}^2}{w^2} +
(1+a^4 w^4)^{1/2}\frac{dw^2}{w^2} \right] 
\eea  

On the other hand, to compare our Informations Metrics (\ref{3.70}) and
(\ref{3.99}) in the previous section with the metric (\ref{4.04}), we
consider the following change of variables, 
\bea\label{4.05}
w= \beta\rho \left(1+\beta_1
\frac{\zeta}{\rho^2}+\beta_2(\frac{\zeta}{\rho^2})^2 + \cdots\right)
\eea
where $\beta$ and  $\beta_i$'s are arbitrary constants. Substituting
(\ref{4.05}) in the metric (\ref{4.04}) will give the following result:
\bea\label{4.10}
ds^2 &=& \frac{\alpha'R^2}{\sqrt{\hat{g}}}\left[\frac{1}{\beta^2\rho^2}
\left(1-\frac{2\beta_1\zeta}{\rho^2} 
-\frac{(-6\beta_1^2+4\beta_2+\lambda^2)\zeta^2}{2\rho^4} +{\cal 
O}(\frac{\zeta}{\rho^2})^3\right)d\vec{x}^2 \right. \nn\\
& & \left. \h1
+\frac{1}{\rho^2}
\left(1-\frac{4\beta_1\zeta}{\rho^2} 
+\frac{(16\beta_1^2-16\beta_2+\lambda^2)\zeta^2}{2\rho^4} 
+{\cal O}(\frac{\zeta}{\rho^2})^3\right)d\rho^2\right]
\eea

Note that $\lambda$ does not appear in the first order of $\zeta$. Indeed,
up to the first order term in $\zeta/\rho^2$, choosing the following
parameter will map both metrics (\ref{4.00}) and (\ref{3.10}) to the 
metric (\ref{3.99}) resulting from Information Metric on the Instanton
Moduli Space:
\bea\label{4.15}
a^{\mu}= \frac{x^{\mu}}{\beta}
\h1 \beta_1=\frac{1}{2} 
\eea
However, note that in this way, it is impossible to map the metric
(\ref{3.70}) to the metric in (\ref{4.04}). So, it seems that only the
modified metric in (\ref{3.99}) is compatible with Supergravity at least,
up to first order in $\zeta$. 

%

\section{Summary and Conclusion}

In the commutative background a map between YM-Instantons and gravity
D-Instantons Moduli can be obtained directly by the Information Metric on
the YM-Instanton Moduli \cite{narain}. We investigated a possible
extension of the Information Metric to the non-commutative instantons.
Firstly, we defined the Metric by simply replacing the instanton density
by its operator version in the NC theory. This Metric can be calculated
perturbatively in the non-commutativity parameter, $\zeta$. In the first
order of $\zeta$, the naive definition of the metric is not in agreement
with the Supergravity result.

We modified the definition by putting a cut-off around the location of
instanton. This cut-off is a ball around the instanton with radius in
order of $\zeta$. Now the modified Information Metric is compatible with
the Maldacena-Russo type solutions in the supergravity
\cite{maldrusso,das}, at least in the first order of $\zeta$. 

The issue of higher order matching can shed more light on this idea that
{\it the (NC) YM-Instanton moduli space is isomorphic to the bulk space in
the context of Gauge/Gravity correspondence}. Note that this
correspondence is different from the ordinary AdS/CFT since a) we are not
using the conventional $L^2$ metric in the field configuration space, even
in the commutative case, and b) this correspondence works for small rank
and small instanton number. 

The next important issue can be the multi-instanton moduli space. It can
be restricted to the simple situation of dilute gas of instantons e. g.
two separated instantons or very close two instantons for the latter case
the commutative moduli is singular but the non-commutative moduli space
can be considered as a resolution of the singularity and the Information
Metric may have an interesting interpretation.

\section*{Acknowledgment}

The author is very grateful to Gautum Mandal and Spenta Wadia for very
fruitful discussions and reading the manuscript. Thanks also to Anindya k.
Biswas, Avinash Dhar, Rajesh Gopakumar and Sandip Trivedi for useful
comments and discussions. This work was partially supported by a travel
grant from the Institute for Studies in Theoretical Physics and
Mathematics, Tehran, Iran.



\end{document}